# Quantitative Electromechanical Atomic Force Microscopy

*Liam Collins[†], Yongtao Liu[†ς], Olga Ovchinnikova[†] and Roger Proksch[‡*]*

[†]Center for Nanophase Materials Sciences, Oak Ridge National Laboratory, Oak Ridge, Tennessee 37831, USA

[ς]Department of Materials Science and Engineering, University of Tennessee, Knoxville, Tennessee 37996, USA

[‡]Asylum Research, an Oxford Instruments Company, Santa Barbara, California 93117, USA

*Corresponding Author: Roger.Proksch@oxinst.com

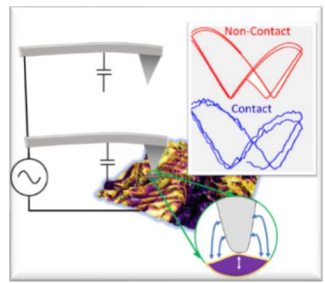

TABLE OF CONTENTS GRAPHIC

ABSTRACT

The ability to probe a materials electromechanical functionality on the nanoscale is critical to applications from energy storage and computing to biology and medicine. Voltage modulated atomic force microscopy (VM-AFM) has become a mainstay characterization tool for investigating these materials due to its unprecedented ability to locally probe electromechanically responsive materials with spatial resolution from microns to nanometers. However, with the wide popularity of VM-AFM techniques such as piezoresponse force microscopy (PFM) and electrochemical strain microscopy (ESM) there has been a rise in reports of nanoscale electromechanical functionality, including hysteresis, in materials that should be incapable of exhibiting piezo- or ferroelectricity. Explanations for the origins of unexpected nanoscale phenomena have included new material properties, surface-mediated polarization changes and/or spatially resolved behavior that is not present in bulk measurements. At the same time, it is well known that VM-AFM measurements are susceptible to numerous forms of crosstalk and, despite efforts within the AFM community, a global approach for eliminating this has remained elusive. In this work, we develop a method for easily demonstrating the presence of hysteretic ("ie, false ferroelectric") long-range interactions between the sample and cantilever body. This method should be easy to implement in any VM-AFM measurement. We then go on to demonstrate fully quantitative and repeatable nanoelectromechanical characterization using an interferometer. These quantitative measurements are critical for a wide range of devices including mems actuators and sensors, memristor, energy storage and memory.

Atomic force microscopy[1] (AFM) uses a cantilever carrying a sharp tip that localizes interactions with a spatial resolution well beyond the optical diffraction limit, in some cases to subatomic lateral resolution. Force- and strain-mediated interactions between the nanoscopic AFM tip and sample are deduced, and in some cases quantified, by measuring the motion of the macroscopic cantilever beam to which the tip is attached. One early and consistently successful application of AFM has been to use a conductive tip to measure localized electromechanical coupling.[2] In the context of this work, we define electromechanical coupling as any material that produces a surface displacement or volume expansion driven by an external electric field. Within this definition, the electromechanical response may arise through diverse phenomena including piezoelectricity, electrostriction or Vegard strain.[3] Voltage-modulated AFM (VM-AFM) is defined here as any force-sensitive technique that operates by placing the AFM tip in contact with the sample surface while the tip-sample bias voltage is periodically modulated. The earliest of these techniques, dubbed piezoresponse force microscopy (PFM), is now 25 years old.[2] In PFM, an oscillating electric field from the tip leads to localized deformations of the sample surface originating from the inverse piezoelectric effect. The piezoelectric strain induced in the sample by the tip voltage translates into motion of the cantilever, which is measured and analyzed with the cantilever detection system. The high resolution of an AFM tip has established PFM as the gold standard for characterization of ferroelectric and piezoelectric materials, not only providing high-resolution domain images but also a plethora of hysteretic and spectroscopic information regarding functional response.[4-6]

The ability to map variations in electromechanical functionality across structural inhomogeneities (e.g., domain walls,[7, 8] grain boundaries[9, 10]) contributed to a rise in popularity of PFM, as well as a broadening of applications far beyond traditional ferro- and piezoelectric materials to fields as diverse as biomaterials[11, 12] and photovoltaics.[13] Meanwhile, a related technique called electrochemical strain microscopy (ESM)[10] was developed and applied to a range of non-piezoelectric, but nevertheless electromechanically active, materials. ESM is based on the detection of localized surface expansion (e.g., Vegard strain) linked with changes in the local concentration of ionic species and/or oxidation states in ionic and mixed ionic-electronic conductors. ESM was first applied to the study of ionic motion in batteries,[10] fuel cell electrodes[14] and oxides;[15, 16] and as with PFM, there is a current demand for applications of ESM for nontraditional applications as well as applications in liquid environments.[17, 18]

Beyond PFM and ESM, there are related techniques[19-22] which can capture similar information, all of which fall under our definition of VM-AFM. Most all VM-AFM modes have evolved with the goal of quantifying local electromechanical deformations. At the same time, it is well established that there are significant opportunities for artifacts and crosstalk in VM-AFM to mask the true underlying material functionality.[23, 24] Even after two decades of incremental improvements, these approaches are still plagued by unwanted spurious background and crosstalk signals that hamper quantitative measurements. The real tip motion, and hence sample displacement of interest, can easily be masked by artifacts or signals involving motion of the macroscopic cantilever body driven by nonlocal electrostatic effects between cantilever body and sample[25-27] and affected by local tip-sample interactions such as topography or contact stiffness changes.[28] In addition, instrumental crosstalk, for example where the tip-sample modulation voltage signal is electronically coupled into the detection electronics, can cause additional artifacts that interact with the artifacts mentioned above.[29, 30] For completeness, in the Supporting Information we provide a detailed list of measurement considerations required for quantitative measurements by PFM or ESM.

The impact of these sources of error become especially pertinent for applications involving materials with relatively weak coupling coefficients (i.e., displacements less than a few tens of picometers). Under such circumstances, the sample driven electromechanical response can be on the order of, or even smaller than, the artificial or crosstalk signals in the measurement itself (see SI). Troublingly, these effects likely contribute to a number of recently reported PFM results of piezoelectricity in materials whose crystallographic symmetry forbids such behavior, as well as reports of ferroelectric-like phenomena (e.g., PFM hysteresis loops) in materials that are non-ferroelectric in the bulk or in cases where size effects are expected to suppress ferroelectricity.[31] Similarly, concerns have been raised regarding the veracity of ESM measurements where the formation of ionic (e.g., $Li^+$) concentration gradients is expected to be too slow to contribute to the ESM signal at the frequencies at which ESM is operated, with some exceptions.[32] This would seem at odds with the surprisingly large electromechanical responses (displacements of hundreds of picometers to a few nanometers) that are often measured by ESM. Overall, it is fair to say that interpretation of ESM response has been largely ambiguous to date, and no artifact-free and universally quantitative method for the evaluation of local parameters has been realized so far.

In this paper, we reveal the true impact of artifacts in PFM/ESM and outline the limits of quantitative VM-AFM as commonly practiced. We start by briefly reviewing artifacts (e.g. topographical crosstalk and electrostatic forces that drive cantilever beam motion) as well as highlighting the role of the cantilever beam dynamics in the optical beam detection (OBD) method used in most traditional AFMs. We demonstrate almost universal hysteretic behaviors measured by VM-AFM across a diverse list of materials (i.e., PZT, soda lime glass, ceria, almond nuts). Using the combined tools of a new, noncontact hysteresis measurement along with a recently developed interferometric displacement sensor (IDS) for the AFM, we reveal the observed hysteresis is entirely the result of nonlocalized interactions between the sample and cantilever body and is not a local phenomenon. Using IDS, we further reveal the propensity for crosstalk in PFM/ESM from other material properties that are not electromechanical in nature. We highlight the scientific relevance of such artifacts through a study of twin domain structure in MAPbI$_3$ and demonstrate that for our samples, the twin domains observed by PFM are not electromechanical in nature (i.e., piezo- or ferroelectric or due to electrochemical strain) and are instead related to local elastic strains. Finally, we use this new method to unambiguously obtain crosstalk-free quantitative values for the effective piezo sensitivity ($d_{eff}$) in X-cut quartz. We show that measurements by IDS are independent of frequency, AFM tip parameters, opening the door for quantitative comparison between measurements and with theory.

**RESULTS AND DISCUSSION**

POTENTIAL ARTIFACTS IN VM-AFM

In VM-AFM modes such as PFM and ESM, a conductive AFM tip is held in contact with the sample while an electrical voltage is applied between the tip and the bottom electrode. The resulting sample vibrations acts as a mechanical drive for the AFM tip (and hence cantilever). Even though the sample property of interest is encoded in the *tip* motion, the vast majority of current AFMs use a position-sensitive photodetector to convert the motion of the *cantilever* into a measured voltage $V_{det}$, as shown in Figure 1a.[33] We refer to this detection scheme as optical beam detection (OBD).[34] Notably, OBD represents an *indirect* measure of the tip displacement, as it is fundamentally an angular measurement of the cantilever motion. An alternative detection approach based on an a hybrid IDS-AFM has recently been demonstrated.[35, 36] A key advantage of the IDS is that it provides a more direct measure of tip displacement than OBD, made possible through the ability to control the IDS detection laser position precisely above the tip position.[35, 36]

In the OBD scheme, the measured signal is roughly proportional to the slope (or bending) of the cantilever.[37, 38] Although presented in various ways, here we will denote a proportionality constant called the inverse optical lever sensitivity *InvOLS*, where the cantilever amplitude at a given frequency $\omega$, $A_\omega$ in meters is related to the measured photodetector voltage amplitude $V_{\omega,det}$ through

$$A_\omega = InvOLS \cdot V_{\omega,det} \qquad (1)$$

$InvOLS$ can be estimated in different ways, the most typical typically by pressing the cantilever against a stiff, noncompliant surface a known distance $\Delta z$, while measuring the associated $\Delta V_{DC,\,det}$. The resulting data are fit to a line, and the slope yields an estimate that assumes $InvOLS \approx \Delta z/\Delta V_{DC,\,det}$. Note that one complication of the OBD technique is that there is a correction in the DC and resonance values of $InvOLS$ that can range from -3% to 9% depending on the spot size and position.[39] In PFM measurements, this force curve calibration procedure then allows determination of the piezo sensitivity $d_{eff}$ (described below). For the IDS, the situation is considerably clearer because the interferometer is a sensor that is both directly dependent on displacement rather than angle and that is calibrated by the wavelength of light. Assuming a spot size $d \ll L$, where $L$ is the cantilever length, it therefore will report an output value $A \approx w(x)$, where $0 \leq x \leq L$ is the interferometer spot position and $w(x)$ is the cantilever displacement along the length $x$. In this work, since $d \approx 1.5$ μm and $L \approx 225$ μm, we have assumed $A = w(x)$.

Following Jesse et al.,[40] we define an "effective" inverse piezo sensitivity $d_{eff}$ by

$$d_{eff} = A_{em}/V_{tip}, \qquad (2)$$

where $V_{tip}$ is the applied voltage and $A_{em}$ is the cantilever amplitude in response to the localized electromechanical surface strain. This sensitivity combines the components of the piezoelectric tensor along the z-axis to describe the resulting response of the PFM cantilever to the applied voltage.[41-43] Note that while PFM and ESM are sensitive to fundamentally different imaging mechanisms (i.e., the inverse piezoelectric effect[44] and Vegard strain[45], respectively), both involve units of length/voltage to describe a linear relationship between surface strain and applied voltage (see Supporting Information for more detail). One of the initial assumptions in PFM has been that $d_{eff} = A_{em}/V_{tip} = A_{1\omega}/V_{tip}$, where $A_{1\omega}$ is the amplitude of the cantilever response measured at the drive frequency, usually measured with a lockin amplifier. As we discuss below, this assumption is often incorrect.

One difficulty in measuring sample displacements by VM-AFM operation is the effect of electrostatic coupling between the sample and the cantilever, which is generally responsible for a background signal at the drive frequency.[22, 46] In VM-AFM, the amplitude response $A_{1\omega}$ of the cantilever at the first harmonic of the AC drive voltage is given by a combination of the localized electromechanical response of the sample ($A_{1\omega,em}$), localized electrostatic interactions between the sample and the cantilever *tip* ($A_{1\omega,el}$) and long-range electrostatics interactions between the sample and the cantilever *body* ($A_{1\omega,nl}$). Indeed, for quantitative measurements further consideration should be given to the detected phase response, considered in detail elsewhere.[40, 47-49] In general, the measured phase is a sum of the excitation phase, a cantilever contribution and an instrumental offset, which can be difficult to separate. In addition, there are similar relationships for lateral components; however, here we will only consider the vertical PFM/ESM response.

Given the popularity of VM-AFM techniques, it may be surprising to find that the quantitative characterization of functional parameters represents an ongoing challenge. Accurate measurement of $d_{eff}$ can be hampered by a host of measurement, environmental/sample and instrumentation factors. Measurement issues include (i) uncertainties in the tip-sample mechanical interface, (ii) uncertainties in the calibration of the mechanical and OBD sensitivity (defined below) of the cantilever and (iii) electrostatic forces acting on the cantilever competing with the piezoelectric actuation by the sample.[40, 50] In addition, environmental factors such as the presence of water layers or adsorbates, as well as sample considerations including dead (i.e., non-electromechanically active) surface layers or competing processes (e.g., ferroionic phenomena) can all complicate quantitative interpretation of PFM/ESM data. Finally, other instrumental background signals can cause crosstalk with the true material response, including mechanical instrumental resonances, frequency dependent electronics and crosstalk. The dangers of exciting instrumental electrical or mechanical resonances in the AFM while making PFM measurements have been elaborated already.[26, 51] Briefly, in a poorly designed excitation system, unwanted electrical couplings in the conductive path to the cantilever can drive the "shake piezo" or couple to the photodetector circuit, leading to apparent cantilever motion indistinguishable from motion originating from the sample electromechanical strain.[29, 40, 48] In the AFM used here, these effects have been effectively eliminated through careful design of the electrical signal routing and shielding.

Next, we consider the intrinsic frequency-dependent behaviors expected for ferroelectric materials and ion conducting materials, respectively and contrast the anticipated material response to the typical response measured by PFM/ESM. As shown schematically in Figure 1b, the resonance of a ferroelectric is very high (hundreds of megahertz to gigahertz), well beyond the operational window of commercial AFMs (typically <10 MHz), indicated by the gray region in Figures 1b-d). In contrast, for frequency dependent ESM response, which is related to electromigration and diffusion kinetics of the ions within the material will be largest at low frequencies (millihertz to kilohertz), as shown in Figure 1c.[32, 45] Above some cut-off frequency ($f_{RC}$), which is governed by the diffuse double layer charging time under the tip[45], the magnitude of the measured response is expected roll off dramatically. At very high modulation frequencies $f_{mod} \gg f_{RC}$, ionic motion, and hence Vegard strains, are expected to become negligible, as the ions cannot diffuse fast enough to the applied voltage (i.e., the ions are in a quasistatic state).[45, 52] To summarize, we expect PFM measurements to be largely frequency independent, whereas for ESM measurements we can expect something more closely resembling a sigmoidal behavior.[45] At the same time, it is well known that the drive frequency of the electrical excitation can have a profound effect on the measured PFM/ESM signal.[53, 54]

The challenge for any PFM/ESM measurement is ensuring good sensitivity to the intrinsic material properties of interest, as well as quantitative extraction of these properties, free from the influences of the cantilever and/or background forces. This requires careful consideration of the operation frequency and the role of the cantilever beam which can have a profound effect on the measured signal.[53, 54] Figure 1d is a visual representation of PFM/ESM amplitude vs. frequency response showing a typical simple harmonic oscillator (SHO) type behavior. The observed features, in or below the AFM operational regime, bear little resemblance to the intrinsic material response of either PFM (Figure 1b) or ESM (Figure 1c) measurements. The measured response originates from the cantilever motion and its associated contact resonances, which are driven by local and nonlocal interactions between the AFM probe (i.e., tip apex, cone and cantilever body) and the sample surface.[55, 56]

In most cases, single-frequency VM-AFM operation has been performed at frequencies of a few hundred kilohertz or less[57], with some exceptions.[58, 59] At these excitation frequencies, well below the contact resonance frequency of the cantilever, interpretation is assumed to be more straightforward to interpret than high-frequency ones.[60,61] Unfortunately, when there are long-range interactions present, this assumption is incorrect. For example, in an earlier study, we found that long-range electrostatic interactions between the body of the cantilever caused cantilever dynamics that led to incorrect phase shifts and significant electrostatically driven amplitudes from the contact resonance all the way down to DC, depending on the positioning of the optical spot on the cantilever (see for example Figure 4 in ref 39). In addition, low-frequency measurements are also more sensitive to 1/$f$ noise, which becomes more significant as the material responsivity gets weaker.

There are potential advantages to operation at higher frequencies, including improved signal-to-noise ratio. Furthermore, higher-frequency measurements are needed for faster scanning, which helps to reduce the impact of 1/$f$ noise and drift and is essential for rapid domain mapping.[58-60] However, complications due to changes in the contact resonance behavior of the AFM cantilever start to play a significant role at high frequencies. Changes in the contact resonance shape as the cantilever scans over the surface can lead to artifacts in the response, or "topographical crosstalk", arising from changes in tip-sample contact area and stiffness (see Ref. 35 for a complete discussion). Crosstalk issues in high-frequency operation have been improved through the implementation of resonance-tracking techniques such as scanning probe resonance image tracking electronics (SPRITE),[62, 63] band excitation (BE)[28, 64, 65] and dual AC resonance tracking (DART).[28] By tracking and characterizing the resonance, it is possible to greatly enhance the measured signal while simultaneously reducing influences from "topographic crosstalk".[66] Resonance tracking techniques allow the determination of the driving force or strain in PFM/ESM by accounting for any change in the contact resonance frequency or quality factor, analyzed in terms of the simple harmonic oscillator (SHO). With few exceptions[32], ESM measurements are largely operated in resonant tracking modes to utilize resonant amplification of the signal. This requirement is likely a consequence of the very small tip displacements that are expected. At the same time, for a pure ionic conductor such as Li$^+$ ion diffusion in lithium aluminum titanium phosphate, the timescales of chemical diffusion and formation of Vegard strains are estimated to be on the order of seconds, too slow to be measured using resonance tracking approaches,[67] bringing into question the origin of signals measured by ESM.

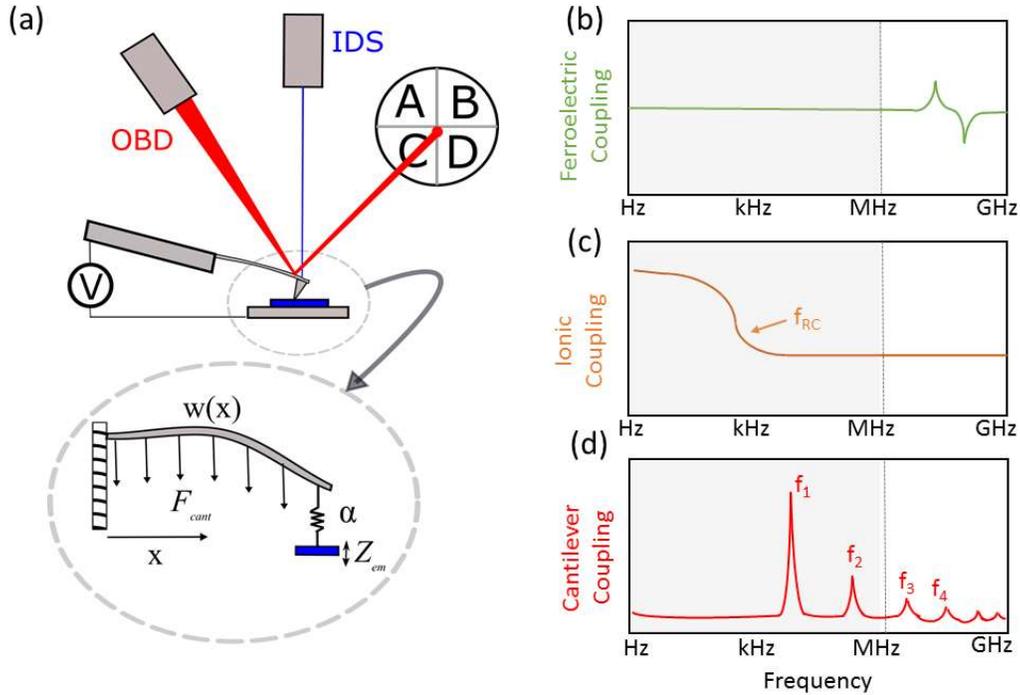

**Figure 1.** (a) Schematic of VM-AFM setup showing both OBD and IDS detection schemes. The cantilever has length $L$, and $w(x)$ is the cantilever vertical displacement at the position $0 \leq x \leq L$. Long-range forces between the sample and the body of the cantilever are denoted by $F_{cant}$, while the localized electromechanical strain $z_{em}$ is coupled to the end of the cantilever ($x = L$) through a normalized tip-sample stiffness α. (b) Qualitative frequency behavior of a ferroelectric material with a flat transfer function below the gigahertz regime, and (c) corresponding behavior of an ionic conductor, in which ion dynamics can easily dominate at lower frequencies (DC to kilohertz frequencies) becoming less pronounced above the cut-off frequency ($f_{RC}$) determined by the diffuse layer charging time. (d) Simplified representation of the measured electromechanical coupling of the cantilever during an out of plane PFM or ESM measurement showing the fundamental contact resonance ($f_1$) along with several higher eigenmodes. The gray regions in (b)-(d) indicate the approximate operational regime of typical commercial AFMs.

In general, quantitative PFM/ESM imaging requires maximizing $A_{em}$ with respect to the other tip or cantilever motions. Unfortunately, electrostatic interactions between the AFM probe and sample can lead to large responses at the drive frequency[23, 40, 68] and hence create background signals that must be either overcome or otherwise eliminated for quantitative PFM measurements.[55, 69] Over the past few years, there has been numerous incremental technological advances that seek to maximize the electromechanical response while minimizing or eliminating the electrostatic components. It has been shown that some artifacts can be reduced by operating with high loads,[68, 70] stiff cantilevers,[40, 68] higher eigenmodes[71], and tall tips.[25] Often, however, the tip-sample stiffness required to eliminate electrostatic effects is sufficiently large to compromises the material, particularly important for fragile thin films or biological materials.[11] Consequently, despite significant efforts, electrostatic interactions remain a significant roadblock towards realizing a widely accepted approach to quantitative VM-AFM.

NONLOCAL HYSTERESIS IN VOLTAGE SPECTROSCOPY PFM/ESM

Next, we investigate the impact electrostatic interactions can have on hysteresis measurements by VM-AFM. Localized hysteresis loops have long been considered strong evidence for nanoscale ferroelectricity[72-79] and ion dynamics in the case of ESM.[64] These loops are typically measured by ramping or stepping a DC voltage, superimposed on a small AC excitation, applied between tip and sample. Switching spectroscopy (SS)[80] is a widely adopted measurement approach that aims to mitigate the effects of electrostatics and is shown schematically in Figure 2a. In SS-PFM or -ESM, the influence of electrostatic forces are reduced by performing remnant measurements between poling steps at zero applied voltage. Importantly, the cantilever is driven by an AC voltage even during the remnant measurement and hence is still subject to electrostatic interactions. As outlined above, these effects are less of a problem when $A_{em} \gg A_{el} + A_{nl}$. However, this condition is rarely met in measurements, especially on highly charged samples, weak ferro- or piezoelectrics and nonpiezoelectric materials such as those explored by ESM.

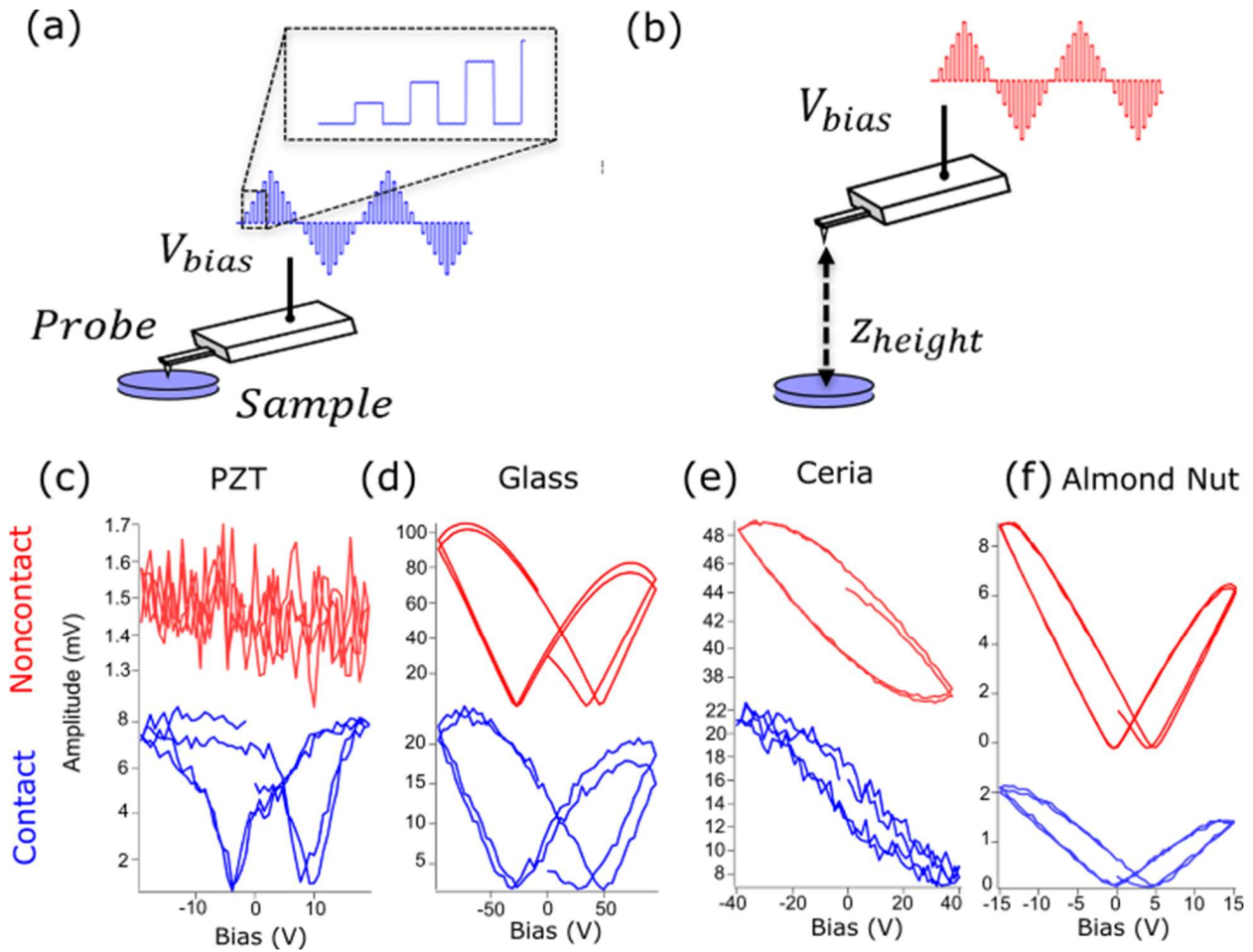

**Figure 2.** Schematics of switching spectroscopy (SS) measurements for (a) contact and (b) noncontact PFM. Remnant hysteresis loops measured on the surface (blue) and 500 nm from the surface (red) on (c) PZT, (d) soda lime glass, (e) ceria, and (f) almond nut. The measurements

were made with an OBD and the amplitudes were left in units of voltage since calibrating the contact optical lever sensitivity is unresolved.

Figure 2 shows amplitude hysteresis loops captured on a series of samples comprising known ferroelectric (lead zirconate titanate, PZT) and ion conducting (ceria and soda lime glass) materials commonly measured by PFM and ESM, respectively. For comparison, we have also included results from a sample with unknown electromechanical behavior (almond nut). Each surface was probed in single-frequency mode using a Pt/Ir coated tip. The modulation frequency (40 kHz) was set to be well below half the contact resonance frequency while care was also taken to avoid the free resonance of the cantilever. Measurements on each sample were performed while the tip was held in contact (bottom row) and out of contact with the sample surface (top row).

When the tip is in contact with the ferroelectric PZT surface, Figure 2c, we observed the expected ferroelectric type switching behavior.[81] When the tip is held in contact with the non-ferroelectric soda lime glass (Figure 2d), we observed hysteretic behavior similar to that previously reported by ESM.[3,82] The observed "elephant ear" shape in hysteresis loops is often attributed to relaxation processes of mobile ions which differentiates these relaxation dynamics from pure ferroelectric polarization switching. To the untrained eye, this hysteretic behavior described could easily be interpreted as ferroelectric switching and highlights the ambiguity in identification of ferroelectricity on unknown materials using VM-AFM techniques.[81] In Figure 2e, the hysteresis loop shape in contact with the sample for ceria differs considerably from that for soda lime glass but again resembles previously reported ESM spectroscopy measurements on ceria.[83] Such behavior was found to match closely with numerical simulations[83] used to describe the local ionic concentration and diffusivity under the tip.

While there have been success describing such hysteretic loops in terms of sample or surface properties, interpretation remains largely ambiguous.[31, 81] Indeed, the misinterpretation of hysteresis loops is not limited to the fields of VM-AFM, and macroscopic polarization-electric field ($P$–$E$) loops are also susceptible to artifacts unassociated with the ferroelectric behavior of the material under test.[84] In a famous work,[84] J. F. Scott demonstrated that $P$-$E$ measurements on ordinary bananas exhibited closed-loop hysteresis nearly identical to hysteresis loops on a true ferroelectric. As a cautionary tale for SS-PFM and -ESM measurements, similar closed-loop hysteresis loops of unknown origin were found to be nearly ubiquitous across the samples tested, even for measurements made in different labs and with different AFM probes and/or operators (not shown), including for an almond nut as shown in Figure 2f. Worryingly, the loops reported here on the non-ferroelectric almond nut bear many of the same characteristics used as indicators for ferroelectricity on materials ranging from perovskite solar cells[85, 86] to aortic walls.[87]

Next, we consider the long-range interactions acting on the cantilever beam and how these influence in the observed hysteretic behaviors. When the measurement on PZT was repeated with the tip held far from the surface, we did not observe hysteretic behavior. The observation of hysteresis loops only when the tip is on contact with the sample would suggest the signal mechanism is mostly electromechanical in nature, as expected for a ferroelectric PZT thin film. Worryingly for any VM-AFM, all materials besides PZT demonstrated similar hysteretic behavior for measurements performed in contact and far away from the surface, even as far as several hundred micrometers from the sample surface (see Figure S1). The observed noncontact hysteresis unequivocally demonstrates that on these samples the measured hysteresis is not due solely to electromechanical strain localized between the tip and sample, as previously believed. Instead, it indicates a signal contribution from long-range interactions between the surface and the body of the cantilever. While measuring long-range hysteretic interactions with conventional OBD AFMs does not yet provide a reliable method for separating long- and short-range effects, the procedure developed herein does provide a simple and universally available means for practitioners to identify the presence of these hitherto difficult-to-understand and -identify artifacts.

A natural question stemming from the observed large long-range hysteretic forces is whether any component of the cantilever motion can be attributed to localized electrochemical strain. To quantify this localized contribution, we used the IDS interferometric method[35, 36] to measure the motion of the cantilever tip separate from the cantilever beam dynamics. While the influence of IDS spot position on PFM imaging contrast of ferroelectrics has previously been reported,[35, 36] here we demonstrate the influence spot position has on SS-PFM/ESM measurements. Figure 3 shows results for soda lime glass when the tip is in contact with the surface. When the IDS laser spot is in front of (Figures 3a and 3b) or behind (Figures 3e and 3f) the tip location, we detect the cantilever contact resonance peaks in the frequency spectra and hysteresis loops similar to that measured by OBD (Figure 3d inset). In direct contrast, when the IDS laser spot is positioned directly over the tip location (Figures 3c and 3d), a frequency-independent response (i.e., no cantilever resonance) and hysteresis free signal is measured. This result is in agreement with previous reports[35, 36] that when the IDS spot is positioned directly over the tip, the detection signal is insensitive to the motion of the cantilever and detects only the displacement of the tip, a prerequisite for quantifying surface strain. This result also compounds the previous conclusion that in many cases, the butterfly loops are not a result of localized surface displacements under the tip; instead, they are an effect of the cantilever motion and the detection scheme, making them inherently sensitive to nonlocal electrostatic interactions acting on the body of the cantilever. It is troubling that these results would seem to indicate that the observed nonlocal interaction can easily dominate the measured response, and ultimately lead to misinterpretation of local material behavior using traditional VM-AFM.[82]

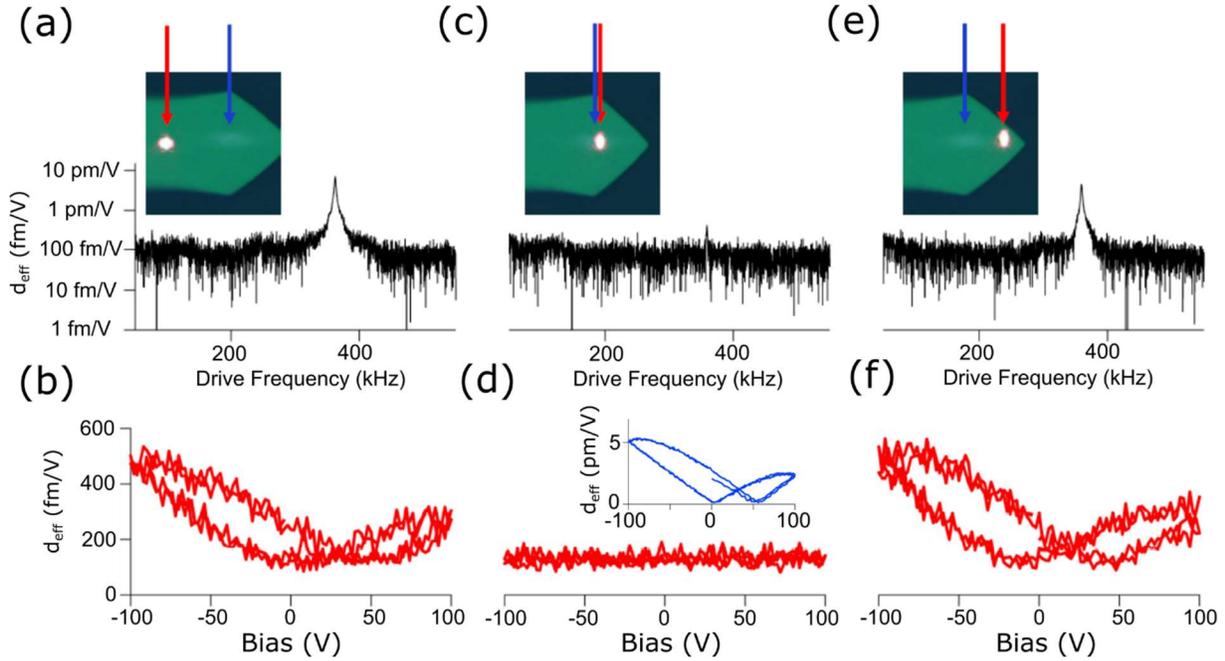

**Figure 3.** Effective $d_{eff}$ measurements on soda lime glass measured by interferometric (IDS) and optical beam (OBD) detection. (a,c,d) $d_{eff}$ as a function of modulation frequency for the three interferometer spot positions. In the optical images of the cantilever, blue arrows indicate the IDS laser spot (just visible), while the red arrows point to the OBD laser spot. (b,d,f) Corresponding hysteresis loops measured withIDS. The inset in (d) shows the $d_{eff}$ hysteresis measured by OBD. The drive frequency was 40 kHz, and the drive amplitude was 3 V. The noise floor in (d) shows that the electromechanical coupling of this glass sample is $d_{eff} \leq 140$ fm/V.

IMAGING ARTIFACTS IN VM-AFM

Considering the results shown in Figures 2 and 3, next we aim to investigate the sensitivity of VM-AFM imaging to local changes in material properties or imaging conditions unrelated to the electromechanical functionality of interest. Importantly an indirect consequence of background forces electrostatically actuating the cantilever beam is that the dynamic actuation can lead to crosstalk with other material properties such as local mechanical properties of the tip-sample junction. As an example of this, we present results in Figure S2 for a polymer composite (polysterene-polycaprolactone, PS-PCL)[88] demonstrate the difficulties in separating true electromechanical response (in presence of background electrostatic forces) from other possible material functionalities (e.g., changes in elastic modulus). Although the PS-PCL test sample is not electromechanically responsive, we observed clear contrast in the DART-PFM amplitude and a ~180º phase inversion between material components. Worryingly, such contrast in amplitude and phase could easily be misinterpreted as electromechanical or piezoelectric behavior. Another example is shown in Figure S3, which demonstrates the propensity for artifacts in ESM imaging arising from changes in tip-sample contact area on ceria, an extensively-studied material. Both examples act as a stark warning for PFM/ESM and related VM-AFM imaging on samples with known weak, or unknown electromechanical responsivity. Furthermore, in light of continued applications on soft materials having heterogenous elastic properties (e.g., biological materials[11, 12], conjugated polymers[17, 18]) these results demonstrate the necessity for more robust and universal imaging approaches which are not sensitive to local changes in elastic modulus.

To demonstrate the immediate scientific relevance of these results, we performed PFM imaging of methylammonium lead triiodide ($CH_3NH_3PbI_3$ or $MAPbI_3$), a hybrid organic-inorganic perovskite (HOIP). HOIPs have achieved great interest in recent years for high-efficiency photovoltaic applications,[89, 90] but many questions remain about the intrinsic properties of these materials. The initial detection of highly-ordered twin domains in $MAPbI_3$ by PFM imaging[91] sparked a rapid rise in applications of PFM on HOIPs, as researchers attempted to unravel the hotly-debated ferroic properties of these systems.[86, 92-100] These efforts have led to seemingly contradictory results claiming both ferroelectric[86, 92-95, 99, 100] and non-ferroelectric[91, 97] behavior.

Many of these PFM studies relied on single-frequency operation, and in almost all of these the drive frequency was close to the cantilever contact resonance frequency,[91] while more recent applications have adopted contact resonance tracking approaches including DART[93, 95] and BE imaging.[98, 101] Presumably the requirement for resonant enhancement stems from a low electrochemical response in this class of materials, although its value has not been reported by PFM so far. At the same time, as discussed above, high-frequency operation necessitates careful consideration of measurement sensitivity to artifacts, even when using resonance tracking techniques that help account for "topographical crosstalk".[56] Table S1 summarizes the mode of operation and other important experimental parameters used in reports of twin domains in HOIPs by PFM.

Figure 4a shows the topography of a region of MAPbI$_3$ with typical micrometer-sized grain structure; while Figures 4c-e show with the corresponding PFM amplitude images acquired using IDS. The images were collected consecutively at the different IDS laser spot positions indicated in Figure 4b. All measurements were captured with an AC voltage of 2 $V_{p-p}$ and a drive frequency of 300 kHz. For all locations except B, twin domains similar to those previously reported by PFM are visible.[91, 92, 99, 100, 102, 103] Interestingly, when the IDS laser spot is located at position B over the AFM tip (Figure 4d), no domains can be observed. This result suggests that the imaging mechanism of the twin domains is different from the expected vertical tip displacement; instead, it is a coupling between sample properties and the cantilever motion. In a recent paper on this topic, we concluded that the observed twin domains were concurrent with variations in elastic, rather than ferroelectric, properties.[101] For comparison, the twin domain structures measured using traditional OBD-based BE-PFM are provided in Figure S4.

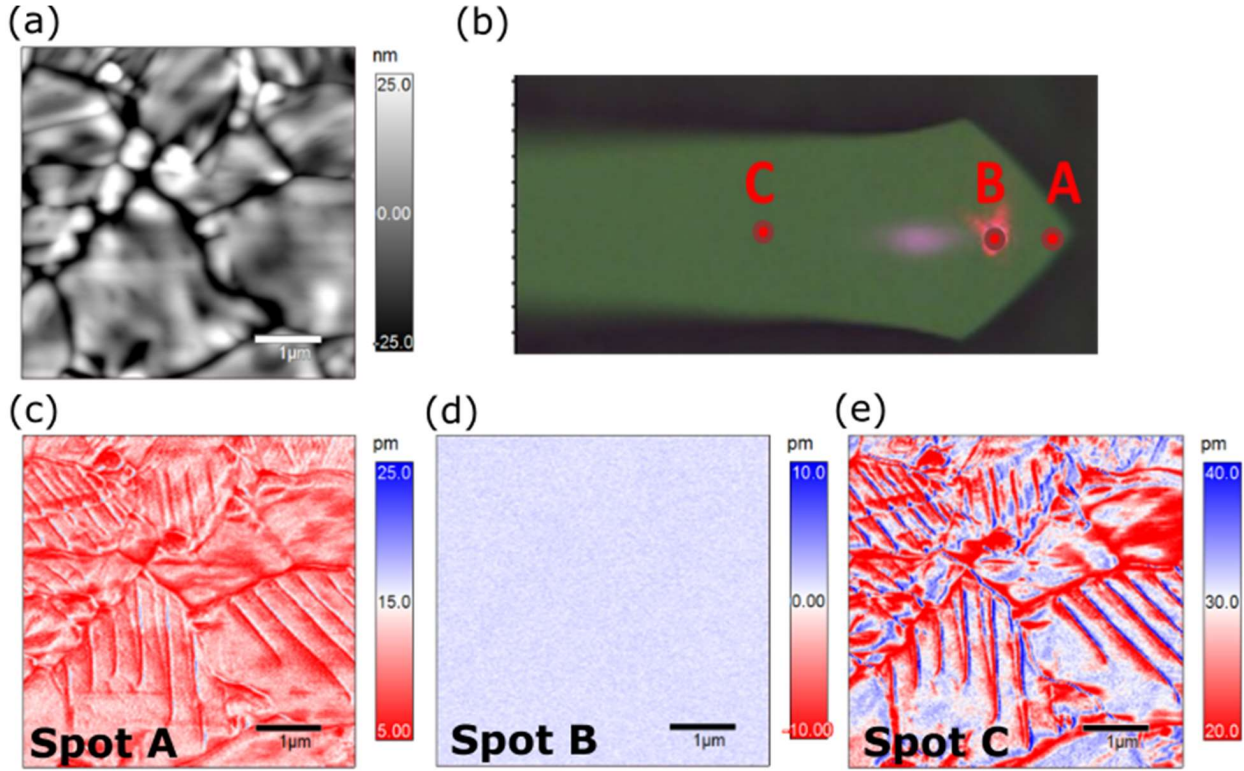

**Figure 4.** (a) AFM topography image of MAPbI$_3$ thin film. (b) Optical image indicating the interferometer spot positions at which PFM imaging was performed. (c-e) PFM amplitude images recorded for the spot positions A-C, respectively. The images were acquired in approximately the same region of the sample.

QUANTITATIVE MEASUREMENTS ON WEAKLY RESPONSIVE MATERIALS

The primary goal of any PFM/ESM measurement should be to accurately measure the voltage-dependent displacement or expansion of a material, which is a fundamental requirement for accurate quantification of the intrinsic $d_{eff}$ of the material. Unfortunately, extraction of quantitative values in PFM or ESM is complicated[60] for the reasons outlined throughout this manuscript. Figure 5, we investigate the limits of quantitative PFM using a piezoelectric X-cut quartz (MTI, size:10x10x0.1 mm, orientation:1120 with edge 0001)) with a relatively low bulk piezoelectric coefficient. Indeed, due to the precisely-known bulk value $d_{33}$ = 2.3 pm/V for the piezoelectric coefficient of bulk X-cut quartz, this sample is sometimes erroneously used as a calibration standard to correct nanoscale PFM measurements on unknown samples.[104] Meanwhile, the universality of such calibration approaches remains questionable as the presence of background forces would erroneously propagate the crosstalk and parasitic signals into further measurements.[26, 60, 61] To the best of our knowledge, only one other set of quantitative PFM measurements on X-cut quartz has been reported. Jungk et al.[61] compared piezoelectric coefficients using both macroscopic top electrodes and a PFM tip. Unfortunately, in the bespoke low-frequency (~5 Hz) PFM setup required for these experiments it took several minutes to collect a single data point, making imaging impossible.

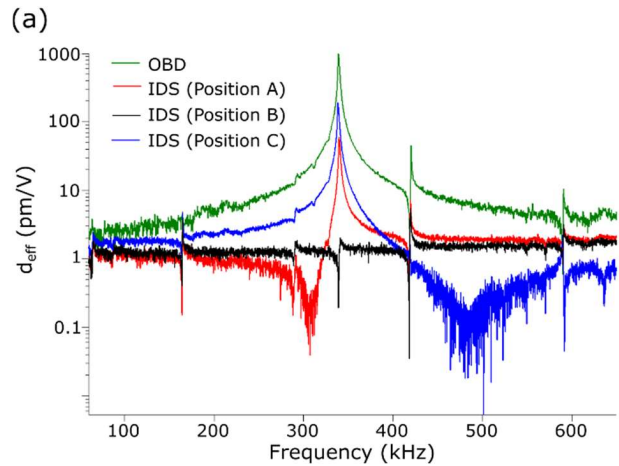

**Figure 5.** Measured $d_{33}$ values measured on x-cut quartz as a function of frequency for OBD (green) and IDS measurements where the detection spot is placed in-front of (red), over (black) and behind (blue) the tip position. The black curve, in the null position indicates $d_{eff} = 1.25$ pm/V.

Figure 5 shows the frequency dependence of the $d_{eff}$ values measured by IDS-PFM on a quartz sample alongside results obtained by OBD detection. The largest frequency excursion was observed for the OBD measurement, which varied by over two orders of magnitude across the 450 kHz measurement window. A strong frequency dependence in $d_{eff}$ was also observed in the IDS measurements when the spot was located away from the tip position (points A and C). In contrast, a frequency-independent value $d_{eff}$ = 1.25 ± 0.1 pm/V was recorded when the IDS spot was located directly over the tip. Notable the measured value by IDS was close to half the expected value from bulk measurements (2.3 pm/V). In Supplementary Figure S5 we investigated the influence of the probe on the measurement for a fresh quartz sample (MTI, size:10x10x0.5 mm, orientation:1120 with edge 0001)) by repeating measurements using a variety of tips having different stiffness, tip coating, and radius as summarized in Table S2. In this case, the measured value of $d_{eff}$ (mean +/- standard deviation) were determined from 30 points across a 20 µm grid. The measured values ranged from (1.45-1.6 pm/V) independent of tip parameters. We did note small variations in measurements performed on different samples or different days which we could attribute to sample condition (e.g. presence of water layer, adsorbates etc.), and independent of measurement of tip parameters. We further noticed that after prolonged exposure of a quartz sample to ambient conditions, the measured values tended to give a further reduction in the coefficient. We confirmed these measurements to be real and due to sample effect by repeating measurements on fresh and aged sample using a variety of AFM probes, all of which gave similar values for $d_{eff}$ that was independent of the cantilever. The data for aged sample is provided in supplementary table S3. Interestingly, for the aged sample the range of values measured across all AFM probes (0.6-0.8 pm/V) matched values reported by Jungk et al (0.8 pm/V).[61] By comparing their low frequency

PFM results with macroscopic measurements using top electrodes they concluded the reduction in coefficient measured locally was due to the inhomogeneous electric field at the tip.[61]

Consistent for all out measurements is the all tips measured a reduced piezoelectric coefficient from bulk values. To check the universality of this observation we repeated measurements for ferroelectric periodically poled lithium niobate (PPLN). This sample represents a good baseline for comparing reproducibility of quantitative methods as it has previously been tested using a variety of methods including IDS-PFM,[36] mode shape correction,[24] and ultra-low frequency PFM.[61] For PPLN, when the IDS spot was placed directly over the tip position we measured a $d_{eff}$ coefficient of ~8.5pm/V (see Supplementary Figure S6). In agreement with measurements on quartz the measured value for LN is well below the expected bulk/macroscopic value (16-23 pm/V), however, it matches precisely with previous measurements using IDS-PFM (8.4 pm/V),[36] and is close to values determined from modal correction of PFM signal (7.5 pm/V)[24] and low frequency KPFM (between 6-7 pm/V).[61] For both samples studied here (PPLN and quartz), IDS-PFM measurements consistently gave piezoelectric coefficients two to three times lower than expected based on reported bulk values. This finding has important implications for quantitative measurements by PFM/ESM as it would suggest that local measurements by a VM-AFM tip might not be directly comparable to macroscopic measurements using top electrodes, which could be due to the inhomogeneities of the electric field.[61] Measurements are ongoing to test this hypothesis and to more closely correlate quantitative measurements of local displacements with macroscopic properties. At the same time, the apparent quantitative agreement between measurements at different frequencies, using different tips, and with previous reports of quantitative PFM,[61] suggests that PFM using IDS represents a universal approach for quantitative PFM, even on samples with low piezoelectric coefficients. Furthermore, the repeatability of the IDS-PFM measurements as shown here demonstrates a universal approach for exploring the bridge between local and macroscopic measurements, or the effect of environmental conditions, free from crosstalk signals and independent of many experimental parameters (e.g. AFM probe) which complicate such investigations using traditional detection methods.

**CONCLUSIONS**

We have shown that bias spectroscopy measurements by PFM/ESM are plagued by nonlocal hysteretic effects that can lead to false claims of ferroelectricity by PFM or may be misinterpreted as localized electrochemical strains in ESM. In the presence of background electrostatics, even qualitative evaluations of PFM/ESM can be hampered by crosstalk with other material functionalities such as elastic modulus. However, the use of IDS-PFM techniques enables decoupling of unwanted cantilever motion from tip displacements. As such, IDS-PFM represents a powerful approach to account for artifacts in PFM/ESM imaging and spectroscopic measurements. Using this method, we discovered that the observed twin domains in MAPbI$_3$ are almost certainly related to elastic strain and further, we have placed a quantitative upper limit on any electromechanical response based on the noise limits of the interferometer. Finally, we have shown that IDS-PFM provides a unique and quantitative measurement of electromechanical coupling coefficients when they are large enough to rise above the instrumental noise background. As such, it presents an obvious opportunity for comparison between experiments performed by different groups, between nanoscopic and macroscopic measurements or between experimental and theoretical results. As such, these measurements represent a paradigm shift in quantitative measurements by VM-AFM.

**EXPERIMENTAL METHODS**

**AFM measurements:** The AFM used in this study combines a commercial Cypher AFM (Asylum Research, Santa Barbara, CA) with an integrated quantitative Laser Doppler Vibrometer (LDV) system (Polytec GmbH, Waldbronn, Germany) to achieve highly sensitive electromechanical imaging and spectroscopy. For Figures 2-5 Pt-coated cantilevers with a spring constant of ~2 N/m and resonance frequency of ~75 kHz were used. All measurements were captured at room temperature under ambient conditions.

**Sample Preparation:** The PZT sample used in Figure 2 was prepared by sol-gel processing. The soda lime float glass was purchase from Fischer scientific cleaned using the procedure described in ref. 83. A raw, unroasted almond nut was to the sample holder and measured without any special sample preparations. Polished x-cut quartz sample was purchased from MTI corporation. All samples were mounted on a steel puck using a small amount of conductive silver paint. Methylammonium lead triiodide film fabrication: Indium tin oxide (ITO) coated glass substrates were sequentially cleaned with deionized water/acetone/isopropanol. The substrates were dried with $N_2$ and treated with UVO before spin-casting perovskite precursors. The perovskite fabrication was conducted in a $N_2$-filled glovebox. The $PbI_2$ (1.2 M in Dimethylformamide) was pre-heated at 100 °C for 10 min and spin-coated onto the ITO glass substrate. After the $PbI_2$ film cooled down to room temperature, the room temperature $CH_3NH_3I$ precursor (0.44 M in ethanol) was spin-coated onto it. The $PbI_2$-$CH_3NH_3I$ bilayer film was annealed at 100 °C for 2 hours to obtain $CH_3NH_3PbI_3$ perovskite.

**AUTHOR CONTRIBUTIONS**

R.P conceived of the log-ranged hysteresis experimental approach. R.P. an L.C carried out acquisition of VM-AFM datasets. Y.L provided the MAPI samples. L.C., Y.L, O.S.O. and R.P.

participated in discussion and interpretation of results. All authors contributed to the writing of the manuscript.

## COMPETING INTERESTS

The authors declare no competing interests.

## ACKNOWLEDGMENTS

(VM-AFM measurements were partially conducted at the Center for Nanophase Materials Sciences, which is a US DOE Office of Science User Facility supported under Contract DE-AC05-00OR22725, (L.C., Y.L, O.S.O). The authors also want to thank Donna Hurley at Lark Scientificfor her valuable discussion about quantitative VM-AFM measurements. Ryan Wagner at Asylum Research pointed out a trick for acquiring the long-range hysteresis data that greatly simplified repeated measurements.

Supplementary information

# Quantitative Electromechanical Atomic Force Microscopy

*Liam Collins†, Yongtao Liu†ζ, Olga Ovchinnikova† and Roger Proksch‡*

† Center for Nanophase Materials Sciences, Oak Ridge National Laboratory, Oak Ridge, Tennessee 37831, USA

ζ Department of Materials Science and Engineering, University of Tennessee, Knoxville, TN, USA.

‡ Asylum Research and Oxford Instruments Company, Santa Barbara, California 93117, USA

**Imaging mechanisms and best practices in PFM and ESM imaging:**

For PFM applied to a piezoelectric or ferroelectric material, the applied bias results in a piezoelectric strain of the material $\delta z = d_{eff}\delta V$, which is translated into motion of the cantilever and measured via the cantilever optical detection system. The measured cantilever motion can yield an effective inverse piezoelectric coupling constant, $d_{eff}$, with units of pm/V by dividing the amplitude of the tip displacement by the amplitude of the tip-sample voltage. The inverse piezo sensitivity is frequently in the range of 1-500pm/V. For ESM, Balke et al. (reference 4) derived a related expression where the factor coupling strain to voltage depends on the lattice poisson ratio $\nu$, the lithium ion diffusion constant $D$, a linear relationship between the applied voltage and chemical potential described by $\eta$, and the effective Vergard constant $\beta$ and the measurement frequency $\omega$: $d_{eff} = 2(1+\nu)\beta \frac{\sqrt{D}}{\eta\sqrt{\omega}}$. Although this expression is considerably more complex than the one for piezoelectric coupling, they both describe a linear relationship between the surface strain and the applied voltage and have the same units of m/V.

**Sensitivity and Detection limits:**

Next we estimate the strain noise limit for VM-AFM measurements, allowing us to estimate the noise limit for PFM/ESM type measurements. For weak (or no) electromechanical responses, the signal to noise of a typical AFM can become the dominant factor. For example, a white noise floor of $40 fm/\sqrt{Hz}$ is typical for a large commercially available cantilever. This implies a noise floor of $\delta z \approx 1.2 pm$ in an imaging bandwidth of 1 kHz. For a 1V drive amplitude, this yields a minimum detectable sensitivity of $d_{eff}^{min} \approx 1.2 pm/V$. While at first blush it appears that $d_{eff}^{min}$ could be lowered by simply applying a larger drive voltage, in many cases including thin films and low dielectric breakdown potentials, larger voltages are precluded because they will be destructive to the sample.

**Quantitative VM-AFM:**

One of the ongoing challenges of PFM, ESM, and related techniques is the accurate characterization of functional parameters. In the case of PFM, the most common functional parameter is the inverse piezo sensitivity; quoted in units of nm/V. Issues with accurate measurement of this parameter can be parametrized to include (a) sample effects and (b) instrument/measurement issues. From a sample perspective, complications in quantifying the intrinsic material property of interest can arise from the tensorial nature of electromechanical coupling, as well as surface effects including surface bound charges, dead or water layers which can result in a voltage drop – or reduce the local electric field under the tip. Other issues can arise due to confinement or clamping effects which can complicate comparison between microscopic measurements using a top electrode and local probe measurements. Other samples properties such as sample roughness can cause further significant complications especially in absence of resonance tracking approaches.

From a measurement standpoint, uncertainties in the tip-sample mechanical interface, uncertainties in the calibration of the mechanical and OBD sensitivity of the cantilever and long range electrostatic forces between the body of the cantilever and tip of the cantilever competing with the piezoelectric actuation. Below we provide a list of experimental "best practices" for achieving quantitative measured by PFM/ESM. These have largely been adapted from a previous publication

(1) Choosing a low drive frequency. While this was indeed confirmed, the definition of "low" depends very strongly on the electrostatic term and may in some cases be well below even a few kHz.

(2) If operating on resonance, which is desirable for improved signal to noise, care must be taken in interpreting the response. Specifically, changes in dissipation will change the quality factor and therefore the gain of the resonance amplifier.

(3) Use of smaller cantilevers to reduce the electrostatic coupling between the tip and sample.

(4) Use of longer tips, thus increasing the distance between the cantilever body and sample, reducing the capacitance.

(5) Shielded probes. These may reduce the capacitance but are also more expensive and not as well developed as conventional cantilevers at this point.

(6) Stiffer cantilevers also will reduce the effect of long range electrostatic forces but may be undesirable for thin films and softer materials, since the high loading force may damage the sample.

(7) Positioning the OBD spot closer to the base of the cantilever can reduce the effect of nodal lines on phase and amplitude (at the cost of a reduction in sensitivity).

(8) As pointed out by others, scanning along the edge of a sample may help minimize these long-range electrical Effects.

**Non-local Hysteresis measurement**

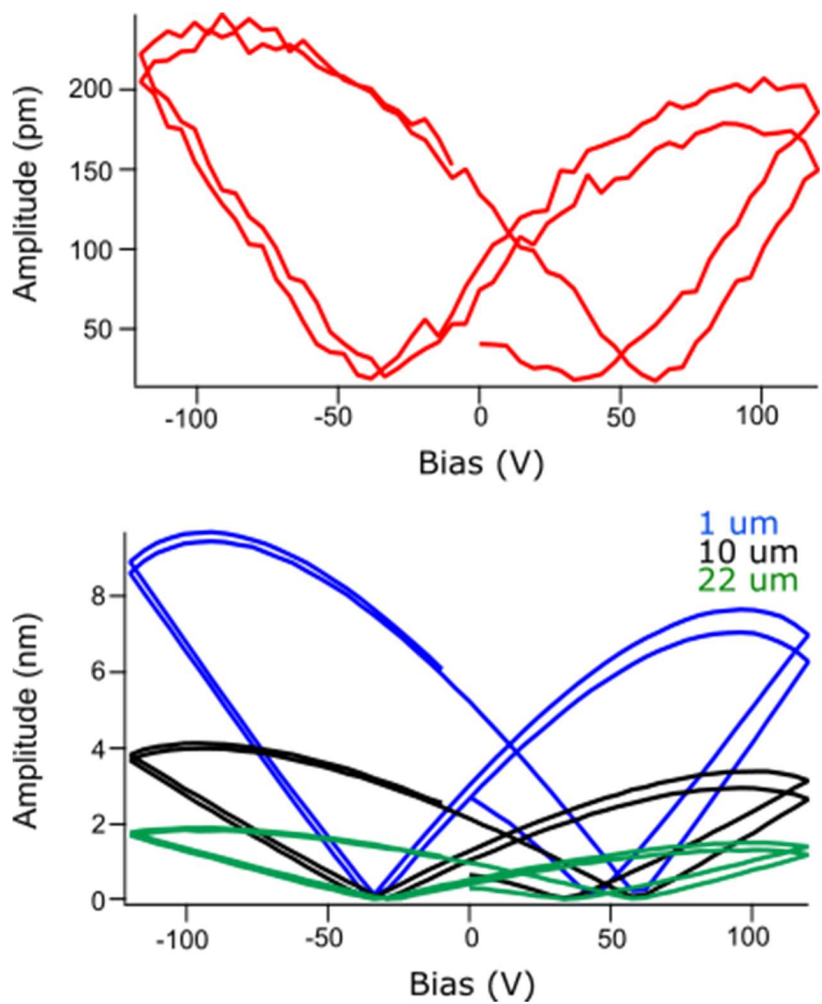

Figure S1: Remnant hysteresis loops measured on the surface (red) and at various heights off the surface of a soda-lime glass sample at various distances from the sample surface. There is a remarkable similarity between the shape of the hysteresis loops, whether collected in contact or out of contact, even the large distances measurements here, strongly suggesting the macroscopic cantilever body governs the measurement ESM response on soda lime glass. As shown in Figure 3 in the manuscript, no surface displacement above the measurements noise floor of $140 fm/V$ c could be detected

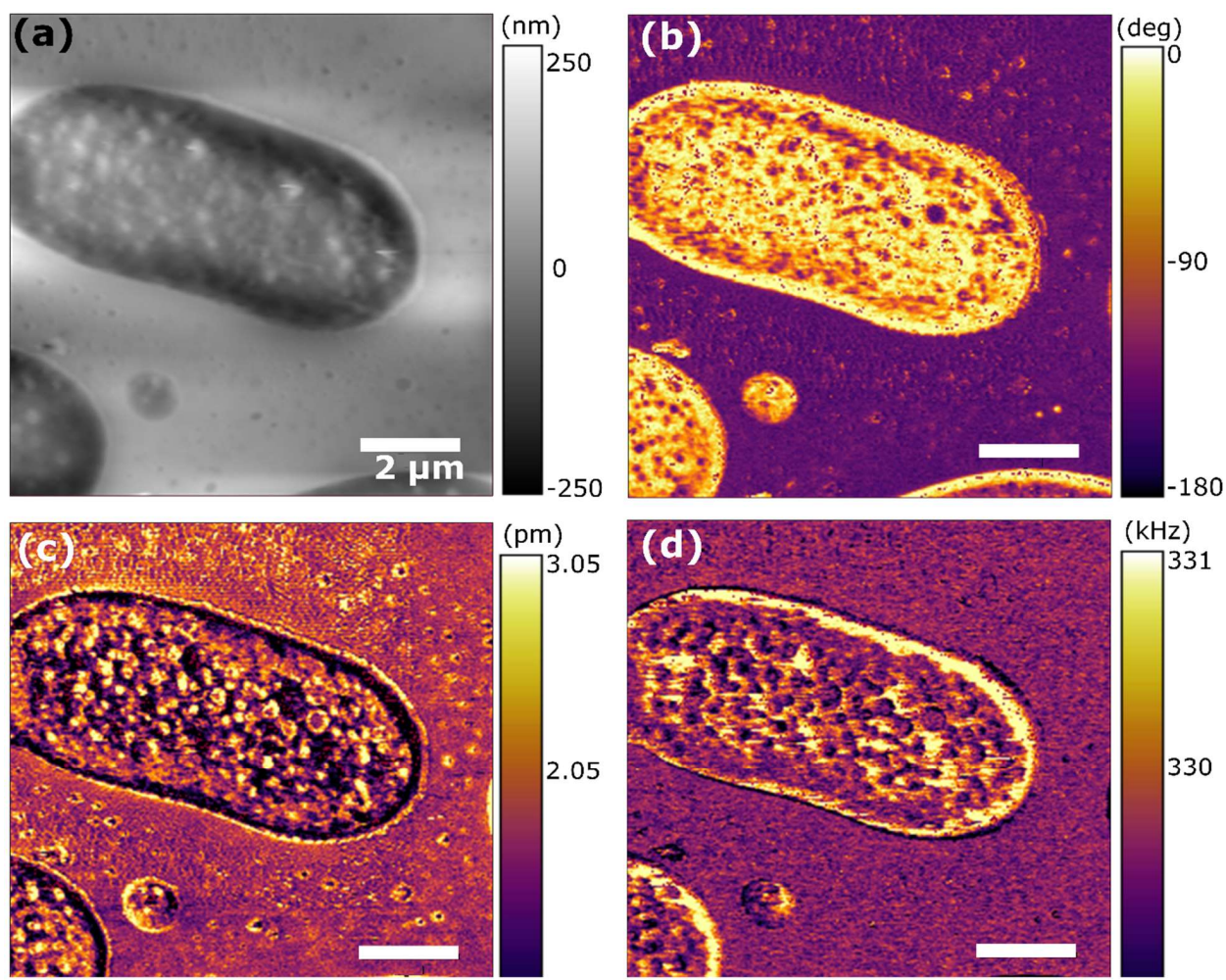

Figure S2: (a) AFM topography image of a polystyrene/polycaprolactone (PS/PCL) blend. (b) PFM phase (c) amplitude and (d) contact resonance measured by DART-PFM.

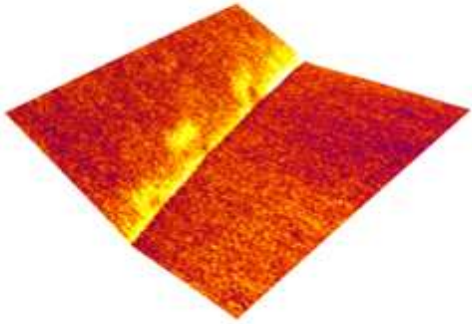

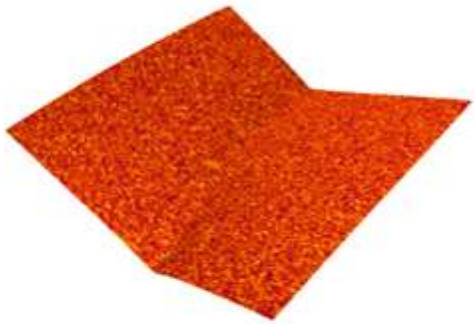

Figure S3: Ceria (a) shows a DART image where amplitude is painted on 3D rendered topography. The enhanced amplitude appears at the grain boundaries. The DART image estimates a $d_{eff}$ of ~10pm/v at the grain boundary (b) shows the same region measured with the interferometer. With this measurement we can put an upper limit of ~1pm/V on $d_{eff}$ for Ceria, implying that the DART/OBD amplitude is instead the result of elastic effects or tip-sample contact area changes at the grain boundary, not electrochemical strain.

Supplementary Table S1: Comparison of imaging PFM parameters used to observe twin domains in MAPbI3

| Publication year | Journal | Material | Technique | Vertical/lateral | Resonance enhancement (single frequency) | Reference |
|---|---|---|---|---|---|---|
| 2016 | *The Journal of Physical Chemistry C* | MAPbI$_3$(Cl) thin film | Single frequency PFM | vertical | Yes | Ref 1 |
| 2017 | *Energy & Environmental Science* | MAPbI$_3$(Cl) thin film | Single frequency PFM | Vertical/lateral | Yes | Ref 2 |
| 2017 | *Science advances* | MAPbI$_3$ thin film | DART PFM | unclear | Yes | Ref 3 |
| 2017 | *ACS applied materials & interfaces* | MAPbI$_3$ thin film | Single frequency PFM and DART | unclear | Yes | Ref 4 |
| 2018 | *ACS Applied Energy Materials* | MAPbI$_3$ thin film | DART | Vertical/lateral | -- | Ref 5 |
| 2018 | *npj Quantum Materials* | MAPbI$_3$ single crystal on TiO2 | DART | vertical | -- | Ref 6 |
| 2018 | *Applied Physics Letters* | MAPbI$_3$ thin film | BE | vertical | -- | Ref 7 |
| 2018 | *Nature materials* | MAPbI$_3$ thin film | Single frequency and BE and LDV | vertical | Yes | Ref 8 |
| 2019 | *Advanced Materials* | MAPbI$_3$ | Single Frequency | vertical | Yes | Ref 9 |

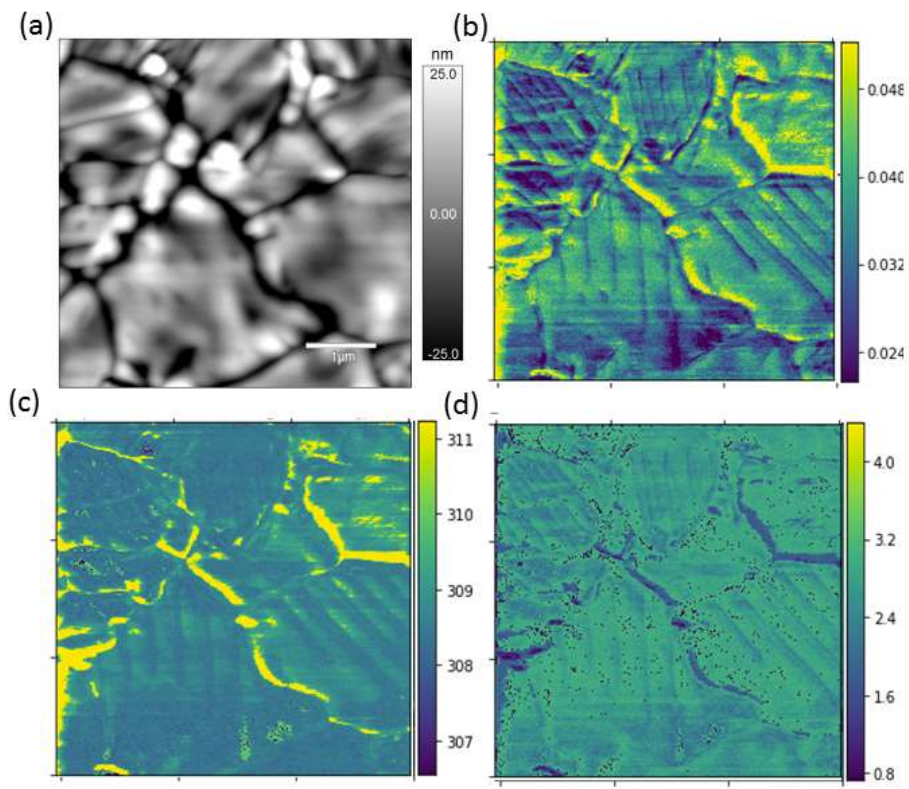

Figure S4: BE- PFM image of MAPbI3 using OBD. (a) Topography, (b) Amplitude, (c) Frequency, (d) Phase determined from SHO fitting on the raw contact resonance amplitude curves.

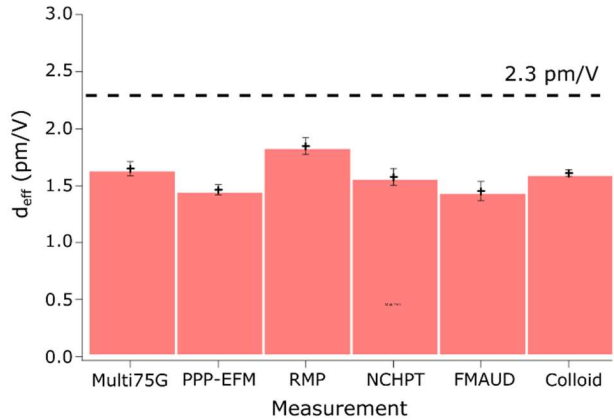

Figure S5: Shows the mean and standard deviation of $d_{eff}$ calculated from 30 frequency sweeps across a 20 um grid using a variety of AFM probes. Probe parameters are described in SI table 2.

**Supplementary Table S2:** Comparison of $d_{eff}$ measurements on a quartz sample using a broad spectrum of different conductive tips. Different Spring constants (2-6 N/m), a variety of metal coating (Pt/Ir, Au, Solid Pt) were investigated.

| Tip | Coating | K (N/m) |
| --- | --- | --- |
| Multi75G (Budget Sensors) | Pt/Ir | 2.9 |
| PPP-EFM (Nanosensors) | Pt/Ir | 2.9 |
| PPP-NCH (Nanosensors) | Pt | 42 |
| PPP-FMAUD (Nanosensors) | Au | 3 |
| Colloid (Nova Scan -30 um) | Au | 6 |
| Rocky Mountain Probe | Pure Platinum | 2 |

**Supplementary Table S3:** Comparison of $d_{eff}$ measurements on an aged (exposed to ambient for several days) quartz sample using different conductive tips.

| Tip | d33 (X-cut Quartz) |
| --- | --- |
| Multi75G (Budget Sensors) | 0.69+0.2 pm/V |
| PPP-NCH (Pt) | 0.6+0.1 pm/V |
| Colloid (Nova Scan -30 um) | 0.79+0.1pm/V |
| Rock Mountain Probe | 0.73 +0.6 pm/V |

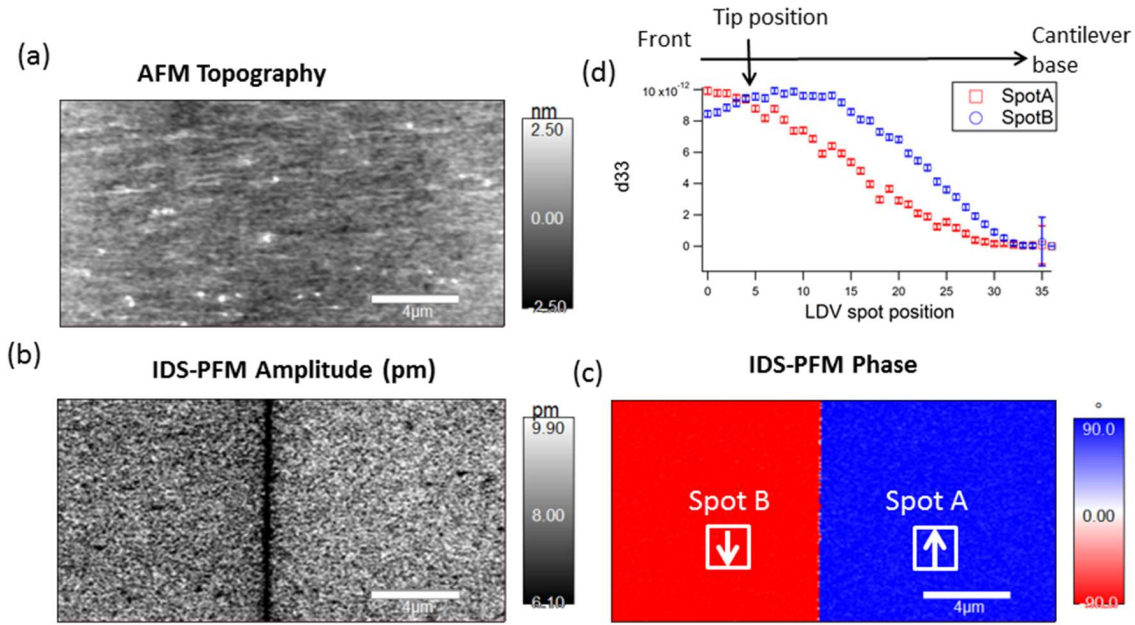

Figure S6: (a) PPLN topography, (b) IDS Amplitude ($d_{33}$ - pm/v; avg ~8.5 pm/V), (c) phase (degree) recorded with the IDS laser spot positioned directly over the tip position on the AFM cantilever. (d) Mode shape spectroscopy, where the Avg (Std) d33 is determined from linear fit of

$V_{ac}$ sweeps, at each position along the cantilever, from the tip to the base in a row of 35 points. Measurement are performed with the tip located at two different domains, spot A and B, demonstrating significant difference in the cantilever motion between up and down polarized domains.